\documentclass{Interspeech2024}
\usepackage{bibspacing}
\usepackage{balance}
\usepackage{tabularx} 

\interspeechcameraready

\title{Genuine-Focused Learning using Mask AutoEncoder for Generalized  Fake Audio Detection}
\small \name[affiliation={1,2}]{Xiaopeng}{Wang}
\small \name[affiliation={1,*}]{Ruibo}{Fu}
\small \name[affiliation={1,}]{Zhengqi}{Wen}
\small \name[affiliation={1,2}]{Zhiyong}{Wang}
\small \name[affiliation={4}]{Yuankun}{Xie}
\small \name[affiliation={2}]{Yukun}{Liu}
\small \name[affiliation={3}]{Jianhua}{Tao}
\small \name[affiliation={1}]{Xuefei}{Liu}
\small \name[affiliation={1}]{Yongwei}{Li}
\small \name[affiliation={1,2}]{Xin}{Qi}
\small \name[affiliation={1,2}]{Yi}{Lu}
\small \name[affiliation={5}]{Shuchen}{Shi}

\address{
  \small $^1$ Institute of Automation, Chinese Academy of Sciences \\
  \small $^2$School of Artificial Intelligence, University of Chinese Academy of Sciences\\
  \small $^3$Department of Automation and Beijing National Research Center for Information Science and Technology, Tsinghua University\\
  \small $^4$ School of Information and Communication Engineering, Communication University of China 
  \small $^5$ Shanghai Polytechnic University
  \small{\thanks{* corresponding author}}
}
\email{wangxiaopeng22@mails.ucas.ac.cn,ruibo.fu@nlpr.ia.ac.cn.}


\begin{document}

\maketitle

\begin{abstract}
    
The generalization of Fake Audio Detection (FAD) is critical due to the emergence of new spoofing techniques. Traditional FAD methods often focus solely on distinguishing between genuine and known spoofed audio. We propose a Genuine-Focused Learning (GFL) framework guided, aiming for highly generalized FAD, called GFL-FAD. This method incorporates a Counterfactual Reasoning Enhanced Representation (CRER) based on audio reconstruction using the Mask AutoEncoder (MAE) architecture to accurately model genuine audio features. To reduce the influence of spoofed audio during training, we introduce a genuine audio reconstruction loss, maintaining the focus on learning genuine data features. In addition, content-related bottleneck (BN) features are extracted from the MAE to supplement the knowledge of the original audio. These BN features are adaptively fused with CRER to further improve robustness. Our method achieves state-of-the-art performance with an EER of 0.25\% on ASVspoof2019 LA.
\end{abstract}
\balance{
\section{Introduction}

Fake Audio Detection (FAD) is vital for ensuring information security and preventing fraud. With the emergence of new spoofing techniques\cite{fu2020focusing}, developing robust FAD measures for Automatic Speaker Verification (ASV) \cite{ASV} systems has become increasingly urgent. FAD systems must have the ability to generalize to adapt to the evolving and unpredictable nature of spoofing attacks in the real world. To promote the development of generalized research on FAD, the ASVspoof \cite{todisco2019asvspoof,wang2020asvspoof} initiative and its series of challenges have amassed a substantial database of genuine and spoofed utterances. Of particular note, the test sets contain numerous spoofing methods not present in the training sets, providing valuable resources for evaluating the performance and generalization challenges of FAD systems.

Currently, the detection of spoofed audio is mainly based on a two-step approach involving feature extraction and classification tasks. The choice of front-end feature extraction is critical for the model's generalizability. Traditional approaches involve manual extraction of features such as first-order spectral features \cite{AASIST, RawGAT, RawGAT-ST, TSSDNet}, second-order spectral features \cite{Graph-ST, MCG-RES2Net50, LCNN_LSTM-sum}, and methods that fuse spectral features \cite{S2pecNet, complex}. These approaches have somewhat improved the robustness of FAD; however, their capability to detect unknown spoofing methods is still limited, and their generalizability needs further enhancement.

In recent years, pre-trained models based on first-order spectral features have demonstrated remarkable noise robustness and generalization capabilities in FAD tasks \cite{xie2023domain, wav2vec2-xlsr_ASP_0.31, wav2vec2_FT, wavLM_FAD}. They use an encoder-only architecture, trained with contrastive loss for the purpose of self-supervised learning \cite{wav2vec2.0, wavLM}. Their outstanding performance is due to pre-training on a large corpus of genuine data with background noise. In contrast, pre-trained models based on second-order spectral features use a different architecture \cite{MAE-ast, audioMAE, ssast}, typically an encoder-decoder structure where the decoder supports the encoder during self-supervised training. However, similar to first-order spectral features, when applied to FAD tasks, these models also function merely as feature extractors, without further modeling the genuine audio patterns. These patterns are present in the FAD training data. In addition, research on better exploiting second-order spectral features pre-training for transfer learning to FAD tasks is still very limited.

In this study, we introduce a novel Genuine-Focused Learning (GFL) framework guided to enhance the generalization capability of FAD, called GFL-FAD. Unlike previous self-supervised models, we adopt an encoder-decoder architecture. To magnify the fine differences between genuine and spoofed audio, a Counterfactual Reasoning Enhanced Representation (CRER) method is proposed in the light of the audio reconstruction mechanism, where Mask AutoEncoder (MAE) \cite{heMAE} architecture is employed for accurately modeling genuine audio features. To reduce the influence of spoofed audio during training, we propose the Genuine Audio Reconstruction Loss (GAR Loss) to maintain the focus on learning genuine audio features, thereby deepening the understanding of the essence of genuine audio and improving the detection ability of the model against unknown spoofing techniques. In addition, content-related bottleneck (BN) features are extracted from MAE to supplement the knowledge of the original audio, and we use an attention mechanism to effectively fuse the BN features of MAE with CRER, which improves the robustness of the model. Evaluated in the ASVspoof2019 Logical Access (LA) scenario, our approach achieves state-of-the-art performance with an Equal Error Rate (EER) of 0.25\%.

\begin{itemize}
\item We propose the GFL-FAD architecture, focusing on genuine audio to amplify the differences between genuine and spoofed audio, thereby enhancing detection capabilities.
\item We propose the GAR Loss to further focus on genuine learning and better capture the distribution of genuine features, thereby improving the generalizability of the model.
\item We use an attention mechanism to effectively fuse the BN features of MAE with CRER, thereby increasing the robustness of the model.
\end{itemize}

}

\begin{figure*}[t!]
  \centering
  \includegraphics[width=\textwidth]{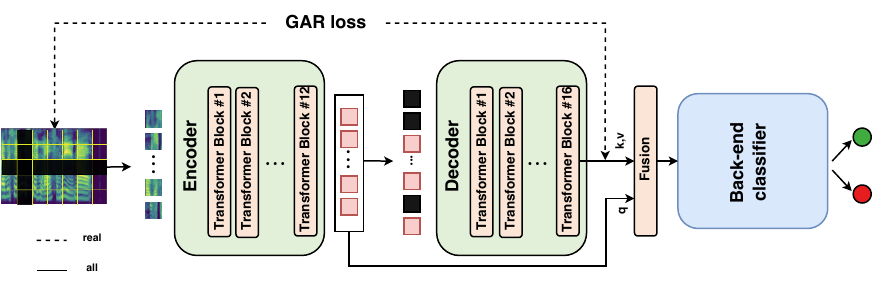} 
    \caption{Illustration of the overall architecture for the proposed GFL-FAD. We start by segmenting patches from the spectral representation. A portion of these patches are encoded by the MAE encoder, resulting in  BN features. The remaining patches are masked and concatenated with the BN features before being fed to the MAE decoder for spectrogram reconstruction. We refer to the reconstructed spectral features as CRER. If the input audio is genuine, an additional GAR Loss is computed to maintain focus on modeling genuine audio knowledge. Through fusion, the CRER is combined with the BN features and passed to the back-end classification network for final detection.}
    \label{fig: Illustration of the overall architecture for the proposed GFL-FAD}
\end{figure*}

\newpage
\section{Proposed method}
The overall framework of our approach is shown in Figure 1. Unlike previous self-supervised models, which are primarily used as feature extractors, we jointly train the decoder of the MAE to enhance the discrimination between genuine and spoofed audio(Section 2.1). To improve the method's focus on genuine, we introduce the GAR Loss (Section 2.2). In addition, to preserve the original information, we use a fusion module to combine the BN features of MAE with CRER (Section 2.3) for the final classification task.

\subsection{Counterfactual Reasoning Enhanced Representation}
The CRER is extracted through the decoder of the MAE. Our MAE architecture is based on the Audio Masked AutoEncoder (audio-MAE)\cite{audioMAE}, with the encoder comprising 12 layers of ViT-Base (ViT-B), and the decoder being a 16 layers transformer that includes a local attention mechanism. The spectrogram input to the MAE must first be preprocessed, including segmented into small patches, each of which is mapped to a \(C\)-dimensional embedding vector. All embedding vectors are represented as \(X = [x_1, x_2, \ldots, x_N] \in \mathbb{R}^{C \times N}\), where \(N=F \times T\) is the number of patches, \(F\) is the number of sub-bands, and \(T\)  is the number of temporal segments. Position embedding is added to each patch to capture its location and the relative positional relationships between patches. The formula is as follows:
\begin{equation}
X^p = X + E_{pos}
\end{equation}

The sequence \(X^p = [x^p_1, x^p_2, \ldots, x^p_N] \in \mathbb{R}^{C \times N}\) consists of patch embedding with integrated positional information, while \(E_{pos} \in \mathbb{R}^{C \times N}\) denotes the positional embedding for all patches. After obtaining \(X^p\), certain areas of time and frequency domains are randomly masked, and the unmasked patches are fed into the encoder to encode the audio representations directly as content-related BN features. The masked patches, along with the encoded audio representations, are then input into the decoder for spectrogram reconstruction. The decoded features serve as the CRER.

\subsection{Genuine Audio Reconstruction Loss}
A key component of our training strategy is the inclusion of the GAR Loss, specifically designed for GFL, to enhance the model's ability to discriminate between genuine and spoofed audio. The formula is as follows:

\begin{equation}
L_{\text{GAR}} = \frac{\sum_{i=1}^{N} \text{mask}_i \cdot L_{\text{recon}, i}}{\sum_{i=1}^{N} \text{mask}_i}
\end{equation}

where $\text{mask}_i$ denotes the authenticity of sample $i$ (label 1 is the genuine sample), and $L_{\text{recon}, i}$ denotes the reconstruction loss for the $i$-th sample. The formula for the reconstruction loss is as follows:

\begin{equation}
L_{\text{recon}, i} = \text{MSE}(\text{prediction}_i, \text{input}_i)
\end{equation}

We use the Mean Squared Error (MSE) as the reconstruction loss, aiming to minimize the MSE between the CRER and the original spectrogram. In the reconstruction loss, we calculate the loss only for the masked patch blocks, rather than the entire spectrogram. Additionally, to better optimize the model, we standardize the reconstructed CRER by its mean and variance.

\subsection{Fusion Strategy for Feature Characteristics}
In our fusion strategy, we employ the encoder segment of the Transformer model \cite{attention} to integrate the BN features from MAE with CRER. We utilize the multi-head attention mechanism of the Transformer, which conducts several self-attention operations concurrently, to highlight the unique characteristics and focal points of each feature set. The attention computation is formulated as follows:
\begin{equation}
\operatorname{Attention}(Q, K, V) = \operatorname{softmax}\left(\frac{Q K^{T}}{\sqrt{d_{k}}}\right) V
\end{equation}

Here, \(d_k\) denotes the dimension of the keys. CRER and BN features are first mapped to the same dimension through two separate linear layers. the embeddings of BN features to serve as queries (\(Q\)), with CRER acting as both keys (\(K\)) and values (\(V\)). This configuration allows for a targeted fusion of features, emphasizing the significant aspects of genuine audio present in the BN features while enhancing them with the nuanced distinctions provided by CRER.

The multi-head attention mechanism generates \(h\) distinct representations of (\(Q, K, V\)), each undergoing a scaled dot-product attention calculation. The outcomes from each head are then concatenated and passed through a feed-forward layer, as described by:
\begin{equation}
\text{MultiHead}(Q, K, V) = \text{Concat}(head_1, \ldots, head_h)W^O
\end{equation}
\begin{equation}
head_i = \text{Attention}(QW_i^Q, KW_i^K, VW_i^V)
\end{equation}

This fusion strategy emphasizes the strengths of each feature set, ensuring that the integrated representation benefits from the detailed resolution of the BN features and the discriminative power of the CRER.

\subsection{Loss Function}
In our framework, the total loss function \(L_{\text{total}}\) is a blend of Cross-Entropy Loss \(L_{\text{CE}}\) for classification and Genuine Audio Reconstruction Loss \(L_{\text{GAR}}\) for audio reconstruction, formulated as:

\begin{equation}
L_{\text{total}} =  L_{\text{CE}} + \alpha \cdot L_{\text{GAR}}
\end{equation}

Here, \(L_{\text{CE}}\) evaluates the model's ability to classify audio as genuine or spoofed, directly impacting the accuracy of fake audio detection. \(L_{\text{GAR}}\) focuses on the reconstruction of genuine audio, ensuring that the model retains the ability to accurately reproduce the characteristics of genuine audio signals. The coefficient \(\alpha\), initially set to 0.01, is a non-learnable parameter that we investigate in ablation experiments to understand its impact on performance. This optimization enhances the model's focus on classification for spoofed audio while maintaining the integrity of genuine audio reconstruction.

\section{Experimental setup}

\subsection{Dataset details}
Our experiments were conducted using the ASVspoof2019 LA dataset, which is recognized for its extensive collection of genuine and spoofed audio samples, produced via advanced text-to-speech and voice conversion techniques. The evaluation set is particularly noted for encompassing a broader range of spoofing attacks, featuring 13 distinct scenarios (A07-A19). This dataset and the distribution of attacks across the partitions are summarized in Table 1. Additionally, acknowledging the influence of initialization on the performance of spoofing detection systems, the results presented are averaged over multiple runs with varied random seeds, ensuring a reliable and consistent performance evaluation. 

\begin{table}[htbp]
  \centering
  \caption{Overview of the ASVspoof2019 LA dataset}
  \label{tab:dataset_overview}
  \begin{tabular}{@{}cccc@{}}
    \toprule
    Partition & genuine & Spoof & Attacks \\ 
    \midrule
    Training & 2,580 & 22,800 & A01 - A06 \\
    Development & 2,548 & 22,296 & A01 - A06 \\
    Evaluation & 7,355 & 63,882 & A07 - A19 \\
    \bottomrule
  \end{tabular}
\end{table}
\subsection{Implementation details}
In our experimental setup, a raw waveform comprising 64,600 frames (approximately 4 seconds) was acquired for analysis. We use a window size of 25 ms with a hop length of 10 ms to transform the waveform into 128 mel-bank features. We initialized our MAE model using the pre-trained parameters of audio-MAE \cite{audioMAE}. We used AASIST \cite{AASIST} as our back-end classifier. Subsequently, our proposed GFL-FAD method was implemented and trained using the PyTorch framework for 100 epochs. The training process was conducted on an NVIDIA RTX 4090 GPU, utilizing a batch size of 16 to efficiently process the data. To optimize the training procedure, we adopted the AdamW \cite{adamW} optimizer with a learning rate of \(5 \times 10^{-6}\). Additionally, a cosine annealing learning rate decay scheme was employed to dynamically adjust the learning rate throughout the training process, ensuring optimal model convergence and performance. 

For the evaluation of our proposed system, we utilized two principal metrics: the EER and the Minimum Tandem Detection Cost Function (min t-DCF) \cite{min-DCF}, with a particular emphasis on the EER for its direct interpretability. The min t-DCF is also reported as it represents the standard for the ASVspoof challenges, providing a comprehensive assessment in conjunction with ASV systems.

\section{Results}

\begin{figure}[b!]
    \centering
    \includegraphics[width=1\linewidth]{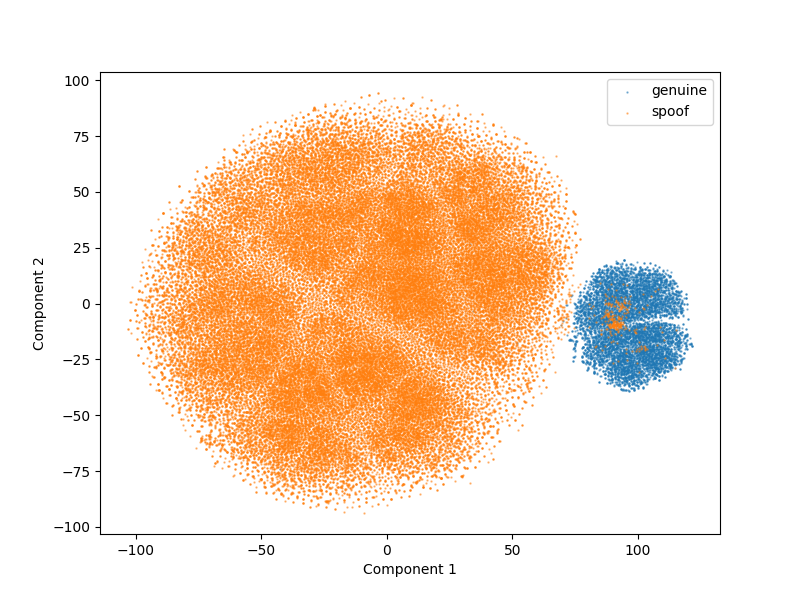}
    \caption{Visualization of High-Dimensional Representations of Genuine and Spoofed Audio Samples using T-SNE
}
    \label{fig:enter-label}
\end{figure}

\begin{table}[t!]
\caption{Performance Comparison of First-Order and Second-Order Spectral Methods With and Without Pre-Training in ASVspoof 2019 LA Evaluation}
\small
\begin{tabular}{lccc}
\toprule
System                 & Front-end    & min      & EER       \\
                       &            & t-DCF       & (\%)     \\
\midrule
  \textbf{No pre-training}                 \\
LCNN-LSTM-sum \cite{LCNN_LSTM-sum}  & LFCC         & 0.0524    & 1.92          \\
Attention+Resnet \cite{Attention_Resnet} & FFT     & 0.0510    & 1.87          \\
MCG-Res2Net50 \cite{MCG-RES2Net50}  & CQT          & 0.0520    & 1.78          \\
FFT-L-SENet  \cite{FFT-L-SENet}     & FFT          & 0.0368    & 1.14          \\
Raw PC-DARTS  \cite{RawPC}          & Raw waveform & 0.0517    & 1.77          \\
Res-TSSDNet  \cite{TSSDNet}         & Raw waveform & 0.0481    & 1.64          \\
RawGAT-ST    \cite{RawGAT-ST}       & Raw waveform & 0.0335    & 1.06          \\
S\(^2\)pecNet \cite{S2pecNet}       & Raw \& LFCC   & 0.0240    & 0.84          \\
AASIST     \cite{AASIST}            & Raw waveform & 0.0275    & 0.83          \\
Graph-ST  \cite{Graph-ST}           & LFB          & 0.0166    & 0.58          \\
DFSincNet \cite{DFSincNet}          & Raw waveform  & 0.0176    & 0.52          \\
F0-SENet34 \cite{F0}                 & STFT         & 0.0143   & 0.43           \\  \hline
\textbf{pre-training} \\
HuRawNet2 \cite{HuRawNet2}                    & Raw waveform  &0.1393     &1.96 \\
Wav2vec2+LGF \cite{wav2vec_Large2_XLSR}         & Raw waveform  & -    &  1.28       \\
Wav2vec2+AASIST  \cite{wav2ve2_AASIST}   & Raw waveform  & -         & 0.90          \\
Wav2vec2+LLGF \cite{wav2vec_Large2_XLSR}         & Raw waveform  & -    &  0.86       \\
Wav2vec2+VIB \cite{wav2vec2_VIB}         & Raw waveform  & 0.0107    & 0.40       \\ 
Wav2vec2+ASP \cite{wav2vec2-xlsr_ASP_0.31}         & Raw waveform  & -    & 0.31       \\ \hline
Ours                   & LFB        & \textbf{0.0071} & \textbf{0.25}          \\
\bottomrule
\end{tabular}
\label{tab:my_label}
\end{table}

\subsection{Performance Comparison}
We visualized the high-dimensional representations of the model at its best performance using T-SNE, as shown in Figure 2.  The visualization confirms the accurate fitting of genuine audio by the model, highlighting the effectiveness of our approach in learning the patterns of genuine data. However, the visualization also shows that some spoofed samples were incorrectly classified as genuine. This may be because the artifacts in the spectra of these spoofed samples are not pronounced, making it difficult for the model to accurately detect them even when the differences are amplified.

In our performance analysis, we listed the performance of SOTA methods based on first-order and second-order spectral features in terms of EER and min t-DCF metrics on the ASVspoof2019 dataset, as shown in Table 2. F0-SENet34 \cite{F0} achieved the best performance by fusing the fundamental frequency with real and imaginary spectrograms in scenarios without pre-training. In scenarios with pre-trained models, the studies were based on first-order spectral features pre-training, with the Wav2vec2+ASP \cite{wav2vec2-xlsr_ASP_0.31} method achieving the best performance. Our GFL-FAD, based on second-order spectral features pre-training, achieved the SOTA performance among all of the above methods. In addition, we compared the performance of our GFL-FAD method with that of using only the encoder model pre-trained with audio-MAE in scenarios using the same classifier, which also demonstrated the effectiveness of our proposed approach. The specific performance details are provided in the ablation experiments.

\subsection{Ablation experiments}

The ablation experiments performed on the ASVspoof2019 LA dataset demonstrate the significant contributions of each component within our GFL-FAD approach, as detailed in Table 3.  Removing the GAR Loss component slightly increases the EER to 0.39\% and the min t-DCF to 0.0115, indicating the importance of GAR Loss within the framework. Furthermore, the absence of the BN features (EN) leads to an increase in the EER to 0.41\%, demonstrating the positive impact of these features on system performance. It is noteworthy that not using the CRER features (DE), which essentially treats the pre-trained model encoder as a mere feature extractor, results in the worst performance. We also explored the optimal setting of the hyperparameter alpha, which balances detection and GAR Loss. The results in Table 4 show that the best performance is achieved when alpha is set to 0.01. In addition, we investigated the effect of different mask ratios (time + frequency) on FAD, as shown in Figure 3. The study revealed that excessively high mask ratios prevent MAE from learning effective audio representations due to a significant loss of spectral information. Conversely, excessively low mask ratios oversimplify the reconstruction task and prevent effective learning by the MAE. Ultimately, the best performance was achieved with a mask ratio of 0.3.

\begin{figure}[!htbp]
    \centering
    \includegraphics[width=1\linewidth]{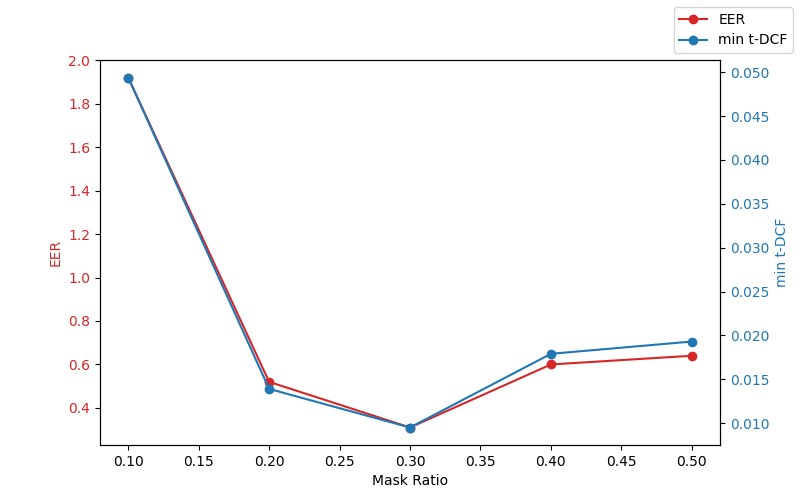}
    \caption{Performance of GFL-FAD at Different Mask Ratios}
    \label{fig:enter-label}
\end{figure}

\begin{table}[!htbp]
  \centering
  \caption{Ablation Experiments of Our Contributions}
  \label{tab:ablation_experiments}
  \scriptsize
  \begin{tabular}{ccc}
    \toprule
    \textbf{Method} & \textbf{EER(\% )} & \textbf{min t-DCF }  \\
    \midrule
    \textbf{GFL-FAD} & \textbf{0.25} & \textbf{0.0071} \\
    w/o GAR          & 0.39  & 0.0115  \\
    w/o EN           & 0.41  & 0.0119\\
    w/o DE           & 0.56  & 0.0184 \\
    \bottomrule
  \end{tabular}
\end{table}

\begin{table}[!htbp]
  \centering
  \caption{Selection of hyperparameter \(\alpha\)}
  \label{tab:hyperparameter_selection}
  \scriptsize
  \begin{tabular}{ccc}
    \toprule
    \textbf{\(\alpha\)} & \textbf{EER(\% )} & \textbf{min t-DCF } \\
    \midrule
    1    & 0.54  & 0.0153 \\
    0.1  & 0.41  & 0.0122 \\
    \textbf{0.01} & \textbf{0.25} & \textbf{0.0071} \\
    \bottomrule
  \end{tabular}
\end{table}

\section{Conclusion}

In this study, we present GFL-FAD, a novel framework that uses GFL to enhance the generalization capabilities of FAD. This approach marks the first application of an encoder-decoder structure, specifically the MAE, to FAD tasks. Our method enhances the subtle differences between real and spoofed audio through the CRER, which is informed by an audio reconstruction mechanism. To minimize the impact of spoofed audio during training, we propose the GAR loss, which ensures focused learning of genuine audio features. This deepens our understanding of the nature of genuine audio and improves the model's ability to detect novel spoofing methods. In addition, we extract BN features from the MAE encoder and fuse them with CRER using an attention mechanism to improve the robustness of the model. Our GFL-FAD demonstrates state-of-the-art performance with an EER of 0.25\% in the challenging ASVspoof2019 LA scenario. In the future, we aim to explore more effective ways to model genuine audio patterns and extend our methodology to other speech anti-spoofing contexts, such as the detection of partially spoofed speech.

\section{Acknowledgements}
This work is supported by the National Natural Science Foundation of China (NSFC) (No. 62101553, No. 62306316, No. U21B20210, No. 62201571).

\bibliographystyle{IEEEtran}
\bibliography{main}

\begin{thebibliography}{10}
\providecommand{\url}[1]{#1}
\csname url@samestyle\endcsname
\providecommand{\newblock}{\relax}
\providecommand{\bibinfo}[2]{#2}
\providecommand{\BIBentrySTDinterwordspacing}{\spaceskip=0pt\relax}
\providecommand{\BIBentryALTinterwordstretchfactor}{4}
\providecommand{\BIBentryALTinterwordspacing}{\spaceskip=\fontdimen2\font plus
\BIBentryALTinterwordstretchfactor\fontdimen3\font minus \fontdimen4\font\relax}
\providecommand{\BIBforeignlanguage}[2]{{%
\expandafter\ifx\csname l@#1\endcsname\relax
\typeout{** WARNING: IEEEtran.bst: No hyphenation pattern has been}%
\typeout{** loaded for the language `#1'. Using the pattern for}%
\typeout{** the default language instead.}%
\else
\language=\csname l@#1\endcsname
\fi
#2}}
\providecommand{\BIBdecl}{\relax}
\BIBdecl

\bibitem{fu2020focusing}
R.~Fu, J.~Tao, Z.~Wen, J.~Yi, and T.~Wang, ``Focusing on attention: prosody transfer and adaptative optimization strategy for multi-speaker end-to-end speech synthesis,'' in \emph{ICASSP 2020-2020 IEEE International Conference on Acoustics, Speech and Signal Processing (ICASSP)}.\hskip 1em plus 0.5em minus 0.4em\relax IEEE, 2020, pp. 6709--6713.

\bibitem{ASV}
D.~A. Reynolds, ``An overview of automatic speaker recognition technology,'' in \emph{2002 IEEE international conference on acoustics, speech, and signal processing}, vol.~4.\hskip 1em plus 0.5em minus 0.4em\relax IEEE, 2002, pp. IV--4072.

\bibitem{todisco2019asvspoof}
M.~Todisco, X.~Wang, V.~Vestman, M.~Sahidullah, H.~Delgado, A.~Nautsch, J.~Yamagishi, N.~Evans, T.~Kinnunen, and K.~A. Lee, ``Asvspoof 2019: Future horizons in spoofed and fake audio detection,'' in \emph{INTERSPEECH 2019-20th Annual Conference of the International Speech Communication Association}, 2019.

\bibitem{wang2020asvspoof}
X.~Wang, J.~Yamagishi, M.~Todisco, H.~Delgado, A.~Nautsch, N.~Evans, M.~Sahidullah, V.~Vestman, T.~Kinnunen, K.~A. Lee \emph{et~al.}, ``Asvspoof 2019: A large-scale public database of synthesized, converted and replayed speech,'' \emph{Computer Speech \& Language}, vol.~64, p. 101114, 2020.

\bibitem{AASIST}
J.-w. Jung, H.-S. Heo, H.~Tak, H.-j. Shim, J.~S. Chung, B.-J. Lee, H.-J. Yu, and N.~Evans, ``Aasist: Audio anti-spoofing using integrated spectro-temporal graph attention networks,'' in \emph{ICASSP 2022-2022 IEEE International Conference on Acoustics, Speech and Signal Processing (ICASSP)}.\hskip 1em plus 0.5em minus 0.4em\relax IEEE, 2022, pp. 6367--6371.

\bibitem{RawGAT}
H.~Tak, J.-W. Jung, J.~Patino, M.~Kamble, M.~Todisco, and N.~Evans, ``End-to-end spectro-temporal graph attention networks for speaker verification anti-spoofing and speech deepfake detection,'' in \emph{ASVSPOOF 2021, Automatic Speaker Verification and Spoofing Countermeasures Challenge}.\hskip 1em plus 0.5em minus 0.4em\relax ISCA, 2021, pp. 1--8.

\bibitem{RawGAT-ST}
H.~Tak, J.~weon Jung, J.~Patino, M.~Kamble, M.~Todisco, and N.~Evans, ``{End-to-end spectro-temporal graph attention networks for speaker verification anti-spoofing and speech deepfake detection},'' in \emph{Proc. 2021 Edition of the Automatic Speaker Verification and Spoofing Countermeasures Challenge}, 2021, pp. 1--8.

\bibitem{TSSDNet}
G.~Hua, A.~B.~J. Teoh, and H.~Zhang, ``Towards end-to-end synthetic speech detection,'' \emph{IEEE Signal Processing Letters}, vol.~28, pp. 1265--1269, 2021.

\bibitem{Graph-ST}
F.~Chen, S.~Deng, T.~Zheng, Y.~He, and J.~Han, ``Graph-based spectro-temporal dependency modeling for anti-spoofing,'' in \emph{ICASSP 2023-2023 IEEE International Conference on Acoustics, Speech and Signal Processing (ICASSP)}.\hskip 1em plus 0.5em minus 0.4em\relax IEEE, 2023, pp. 1--5.

\bibitem{MCG-RES2Net50}
X.~Li, X.~Wu, H.~Lu, X.~Liu, and H.~Meng, ``{Channel-Wise Gated Res2Net: Towards Robust Detection of Synthetic Speech Attacks},'' in \emph{Proc. Interspeech 2021}, 2021, pp. 4314--4318.

\bibitem{LCNN_LSTM-sum}
X.~Wang and J.~Yamagishi, ``A comparative study on recent neural spoofing countermeasures for synthetic speech detection,'' \emph{Interspeech 2021}, 2021.

\bibitem{S2pecNet}
P.~Wen, K.~Hu, W.~Yue, S.~Zhang, W.~Zhou, and Z.~Wang, ``{Robust Audio Anti-Spoofing with Fusion-Reconstruction Learning on Multi-Order Spectrograms},'' in \emph{Proc. INTERSPEECH 2023}, 2023, pp. 271--275.

\bibitem{complex}
C.~Fan, J.~Xue, S.~Dong, M.~Ding, J.~Yi, J.~Li, and Z.~Lv, ``Subband fusion of complex spectrogram for fake speech detection,'' \emph{Speech Communication}, vol. 155, p. 102988, 2023.

\bibitem{xie2023domain}
Y.~Xie, H.~Cheng, Y.~Wang, and L.~Ye, ``Domain generalization via aggregation and separation for audio deepfake detection,'' \emph{IEEE Transactions on Information Forensics and Security}, 2023.

\bibitem{wav2vec2-xlsr_ASP_0.31}
J.~W. Lee, E.~Kim, J.~Koo, and K.~Lee, ``{Representation Selective Self-distillation and wav2vec 2.0 Feature Exploration for Spoof-aware Speaker Verification},'' in \emph{Proc. Interspeech 2022}, 2022, pp. 2898--2902.

\bibitem{wav2vec2_FT}
Y.~Xie, H.~Cheng, Y.~Wang, and L.~Ye, ``Learning a self-supervised domain-invariant feature representation for generalized audio deepfake detection,'' in \emph{Proc. INTERSPEECH}, vol. 2023, 2023, pp. 2808--2812.

\bibitem{wavLM_FAD}
Z.~Cai and M.~Li, ``Integrating frame-level boundary detection and deepfake detection for locating manipulated regions in partially spoofed audio forgery attacks,'' \emph{Computer Speech \& Language}, vol.~85, p. 101597, 2024.

\bibitem{wav2vec2.0}
A.~Baevski, Y.~Zhou, A.~Mohamed, and M.~Auli, ``wav2vec 2.0: A framework for self-supervised learning of speech representations,'' \emph{Advances in neural information processing systems}, vol.~33, pp. 12\,449--12\,460, 2020.

\bibitem{wavLM}
S.~Chen, C.~Wang, Z.~Chen, Y.~Wu, S.~Liu, Z.~Chen, J.~Li, N.~Kanda, T.~Yoshioka, X.~Xiao \emph{et~al.}, ``Wavlm: Large-scale self-supervised pre-training for full stack speech processing,'' \emph{IEEE Journal of Selected Topics in Signal Processing}, vol.~16, no.~6, pp. 1505--1518, 2022.

\bibitem{MAE-ast}
A.~Baade, P.~Peng, and D.~Harwath, ``{MAE-AST: Masked Autoencoding Audio Spectrogram Transformer},'' in \emph{Proc. Interspeech 2022}, 2022, pp. 2438--2442.

\bibitem{audioMAE}
P.-Y. Huang, H.~Xu, J.~Li, A.~Baevski, M.~Auli, W.~Galuba, F.~Metze, and C.~Feichtenhofer, ``Masked autoencoders that listen,'' \emph{Advances in Neural Information Processing Systems}, vol.~35, pp. 28\,708--28\,720, 2022.

\bibitem{ssast}
Y.~Gong, C.-I. Lai, Y.-A. Chung, and J.~Glass, ``Ssast: Self-supervised audio spectrogram transformer,'' in \emph{Proceedings of the AAAI Conference on Artificial Intelligence}, vol.~36, no.~10, 2022, pp. 10\,699--10\,709.

\bibitem{heMAE}
K.~He, X.~Chen, S.~Xie, Y.~Li, P.~Doll{\'a}r, and R.~Girshick, ``Masked autoencoders are scalable vision learners,'' in \emph{Proceedings of the IEEE/CVF conference on computer vision and pattern recognition}, 2022, pp. 16\,000--16\,009.

\bibitem{attention}
A.~Vaswani, N.~Shazeer, N.~Parmar, J.~Uszkoreit, L.~Jones, A.~N. Gomez, {\L}.~Kaiser, and I.~Polosukhin, ``Attention is all you need,'' \emph{Advances in neural information processing systems}, vol.~30, 2017.

\bibitem{adamW}
I.~Loshchilov and F.~Hutter, ``Decoupled weight decay regularization,'' in \emph{International Conference on Learning Representations}, 2018.

\bibitem{min-DCF}
T.~Kinnunen, H.~Delgado, N.~Evans, K.~A. Lee, V.~Vestman, A.~Nautsch, M.~Todisco, X.~Wang, M.~Sahidullah, J.~Yamagishi \emph{et~al.}, ``Tandem assessment of spoofing countermeasures and automatic speaker verification: Fundamentals,'' \emph{IEEE/ACM Transactions on Audio, Speech, and Language Processing}, vol.~28, pp. 2195--2210, 2020.

\bibitem{Attention_Resnet}
H.~Ling, L.~Huang, J.~Huang, B.~Zhang, and P.~Li, ``Attention-based convolutional neural network for asv spoofing detection.'' in \emph{Interspeech}, 2021, pp. 4289--4293.

\bibitem{FFT-L-SENet}
J.~Xue, C.~Fan, Z.~Lv, J.~Tao, J.~Yi, C.~Zheng, Z.~Wen, M.~Yuan, and S.~Shao, ``Audio deepfake detection based on a combination of f0 information and real plus imaginary spectrogram features,'' in \emph{Proceedings of the 1st International Workshop on Deepfake Detection for Audio Multimedia}, 2022, pp. 19--26.

\bibitem{RawPC}
W.~Ge, J.~Patino, M.~Todisco, and N.~Evans, ``Raw differentiable architecture search for speech deepfake and spoofing detection,'' in \emph{ASVSPOOF 2021, Automatic Speaker Verification and Spoofing Countermeasures Challenge}.\hskip 1em plus 0.5em minus 0.4em\relax ISCA, 2021, pp. 22--28.

\bibitem{DFSincNet}
B.~Huang, S.~Cui, J.~Huang, and X.~Kang, ``Discriminative frequency information learning for end-to-end speech anti-spoofing,'' \emph{IEEE Signal Processing Letters}, vol.~30, pp. 185--189, 2023.

\bibitem{F0}
J.~Xue, C.~Fan, Z.~Lv, J.~Tao, J.~Yi, C.~Zheng, Z.~Wen, M.~Yuan, and S.~Shao, ``Audio deepfake detection based on a combination of f0 information and real plus imaginary spectrogram features,'' in \emph{Proceedings of the 1st International Workshop on Deepfake Detection for Audio Multimedia}, 2022, pp. 19--26.

\bibitem{HuRawNet2}
L.~Li, T.~Lu, X.~Ma, M.~Yuan, and D.~Wan, ``Voice deepfake detection using the self-supervised pre-training model hubert,'' \emph{Applied Sciences}, vol.~13, no.~14, p. 8488, 2023.

\bibitem{wav2vec_Large2_XLSR}
X.~Wang and J.~Yamagishi, ``{Investigating Self-Supervised Front Ends for Speech Spoofing Countermeasures},'' in \emph{Proc. The Speaker and Language Recognition Workshop (Odyssey 2022)}, 2022, pp. 100--106.

\bibitem{wav2ve2_AASIST}
H.~Tak, M.~Todisco, X.~Wang, J.~weon Jung, J.~Yamagishi, and N.~Evans, ``Automatic speaker verification spoofing and deepfake detection using wav2vec 2.0 and data augmentation,'' in \emph{Proc. The Speaker and Language Recognition Workshop (Odyssey 2022)}, 2022, pp. 112--119.

\bibitem{wav2vec2_VIB}
Y.~Eom, Y.~Lee, J.~S. Um, and H.~R. Kim, ``{Anti-Spoofing Using Transfer Learning with Variational Information Bottleneck},'' in \emph{Proc. Interspeech 2022}, 2022, pp. 3568--3572.

\end{thebibliography}

\end{document}